\documentclass[
    ,final            
  ]
  {aipproc}

\layoutstyle{6x9}

\begin{document}

\title{Neutrino Mixing in Unified Extended Seesaw Model}


\author{Ivo de Medeiros Varzielas}{
  address={CFTP, Instituto Superior T\'{e}cnico, Av. Rovisco Pais, 1, 1049-001 Lisboa, Portugal}
}



\begin{abstract}
The seesaw mechanism can play a key role in the generation of the leptonic mixing in unified models. We consider an unified model with a family symmetry and extended seesaw, and obtain viable fermion masses and mixing (leptonic mixing is close to tri-bi-maximal).
\end{abstract}

\maketitle


\section{Introduction}

This proceedings submission corresponds to a parallel talk in the SUSY09 conference 
\footnote{
\url{http://nuweb.neu.edu/susy09/ParallelSchedule.php}
} and is based partly on \cite{Ivo7}.


Experimental data \cite{PDG} tells us that the neutrino masses are small compared to the charged fermions, and that lepton mixing features large angles compared to quark mixing - neutrino data \cite{Mariam_data} is consistent with near tri-bi-maximal (TB)
mixing \cite{TBM}. These striking features of the neutrinos may be understood in extensions of the Standard Model (SM). Here we combine a family symmetry (FS) with seesaw mechanisms.

Seesaw mechanisms can provide an elegant explanation for the smallness of neutrino masses. Type I seesaw arises in SM extensions where right-handed (RH) neutrinos are added - large RH Majorana masses lead to small left-handed (LH) Majorana masses \cite{Minkowski:1977sc, Yanagida:1979as, Glashow:1979nm, Mohapatra:1979ia, GellMann:1980vs, Schechter:1980gr}. Type II arises from adding a Higgs isotriplet \cite{Lazarides:1980nt, Mohapatra:1980yp}. There are also extended seesaw mechanisms requiring other extensions of the field content that can be interesting in the context of Grand Unifying Theories (GUTs), such as the linear seesaw \cite{Malinsky:2005bi}.


\section{Tri-bi-maximal mixing through seesaw}


We consider a framework with an underlying $SU(3)_F$ FS (the FS may be a discrete subgroup of it) and underlying $SO(10)$ GUT. The FS is broken by the vacuum expectation values (VEVs) of extra scalars called familons - GUT singlets transforming as anti-triplets of the FS,  denoted $\phi_{A}^i$ ($A$ is a label that identifies which VEV the respective familon acquires, the superscript is the anti-triplet family index). The SM fermions are unified into $16$s of $SO(10)$ and transform as triplets of the FS - the LH fermions denoted $\psi_i$ and the conjugates of the RH fermions denoted $\psi^c_i$ ($c$ is a label, the subscript is the triplet family index). The Higgs are FS singlet $10$s of $SO(10)$ denoted as $H$. In this setup Yukawa terms are no longer constructed from $\psi \psi^c H$, but instead from $\phi^i_{A} \psi_i \phi^j_{B} \psi^c_j H$. The fermion masses arise from the familon VEVs through the Froggatt-Nielsen (FN) mechanism \cite{fnielsen}.

The familons $\phi_{23}$ and $\phi_{123}$ are important ingredients. They acquire orthogonal VEVs:

\begin{equation}
\langle \phi_{23} \rangle = (0,b,-b) \quad ; \quad \langle \phi_{123} \rangle = (c,c,c).
\label{VEVs}
\end{equation}
The VEVs may be complex but must be orthogonal. The $23$ and $123$ labels identify the non-zero entries (which have equal magnitudes within the same VEV). We do not discuss here how to align these VEVs (refer to \cite{Ivo7}, and previous models \cite{Ivo1, Ivo2, Ivo3, Ivo6}).


\subsection{Terms and diagrams}

In order to obtain TB neutrino mixing, the desired effective superpotential is:

\begin{equation}
P_\nu = \lambda_@ \phi^i_{23} \nu_i \phi^j_{23} \nu_j H H + \lambda_\odot \phi^i_{123} \nu_i \phi^j_{123} \nu_j H H.
\label{eff}
\end{equation}
$\nu_i$ are LH neutrinos (FS triplet, belongs to $\psi_i$). With the VEVs in \eqref{VEVs} these terms lead to an atmospheric ($@$) mass eigenstate with equal parts of the second and third component of $\nu_i$ whereas the solar ($\odot$) mass eigenstate is made of equal parts of all three components of $\nu_i$.
The effective superpotential must not have a mixed term $\lambda_{m} \phi^i_{23} \nu_i \phi^j_{123} \nu_j H H$ so we must assume there is some effective symmetry, such as a $Z_2$ under which only one of the familons transforms (e.g. $\phi_{23} \rightarrow - \phi_{23}$, $\phi_{123} \rightarrow \phi_{123}$).

The objective is to obtain \eqref{eff} within a viable GUT. This is rather difficult as the underlying $SO(10)$ is very constraining. In \cite{Ivo1, Ivo3} the type I seesaw mechanism is used to achieve exact neutrino TB mixing. Lepton mixing deviates slightly from TB by receiving small corrections from the charged leptons. The phenomenologically viable models in \cite{Ivo1, Ivo3} can fit fit the charged fermion masses and quark mixing, and the interplay of the neutrino Dirac and Majorana masses leads to TB mixing for the effective neutrinos. Here we focus solely on the neutrino terms, which are:

\begin{equation}
P_Y = \phi^i_{23} \nu_i \phi^j_{123} \nu^c_j H + \phi^i_{123} \nu_i \phi^j_{23} \nu^c_j H .
\label{PY}
\end{equation}
\begin{equation}
P_M = \phi^i_{123} \nu^c_i \phi^j_{123} \nu^c_j (...) +  \phi^i_{23} \nu^c_i \phi^j_{23} \nu^c_j (...).
\label{PM}
\end{equation}
We do not explicitly show the FN messengers masses in the effective terms. The FS triplet $\nu^c_i$ in $P_M$ denotes the conjugate of the RH neutrinos (and belongs to $\psi^c_i$). The $(...)$ stand for a suitable combination of fields that ensure the terms are invariant under the GUT and any auxiliary symmetries - they do not change the structure, only the magnitude of the respective contribution to the Majorana mass.

\begin{figure}
\label{fig:step3}
  \includegraphics[height=.15\textheight]{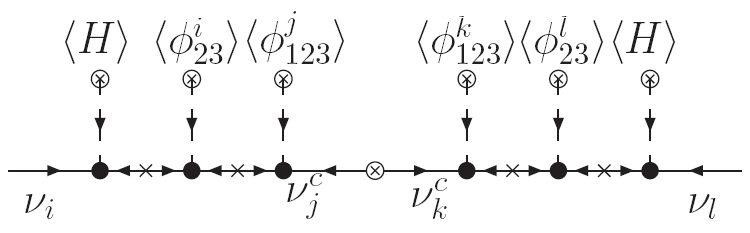}
  \caption{Seesaw diagram corresponding to the atmospheric eigenstate.}
\end{figure}

Possibly the simplest way to understand how this combination of Yukawa and Majorana terms works to produce TB neutrino mixing is by considering the seesaw diagrams expanded to include the familon VEV insertions. This is described in detail in \cite{Ivo5}. For example, in order to obtain the $@$ term of \eqref{eff} we need the family invariant contraction $\phi^i_{23} \nu_i$. Only the first term in
\eqref{PY} has it.
To complete the $@$ term, we need another $\phi^i_{23} \nu_i$. We must get it through some seesaw mechanism to respect the SM gauge group. In this case we can only go through type I seesaw using the first Majorana term in
\eqref{PM} as we started from the Yukawa $\phi^i_{23} \nu_i \phi^j_{123} \nu^c_j H$, which has
$\phi^i_{123} \nu^c_i$. Due to the orthogonality of the VEVs the other term will not contribute. The only Majorana term involved is $\phi^j_{123} \nu^c_j \phi^k_{123} \nu^c_k (...)$, and it repeats its family invariant contraction. In order to complete the seesaw we are once again forced to a single choice by the orthogonal VEVs - we need an Yukawa term with the contraction
$\phi^k_{123} \nu^c_k$. We conclude that this can only be the first term again, which has the desired contraction $\phi^l_{23} \nu_l$. Figure \ref{fig:step3} shows the seesaw diagram with familon insertions (FN messengers are not explicitly shown).

In the original models \cite{Ivo1, Ivo3} the solar eigenstate follows from exactly the same diagram where the familons are swapped (i.e. using only the second terms of \eqref{PY} and \eqref{PM} instead of using only the first terms).

We will now discuss a distinct model based on this framework. The model in \cite{Ivo7} further enlarges the particle content by including $SO(10)$ singlets $s_i$ and $\bar{s}^i$ (FS triplet and anti-triplet respectively). These allow extra freedom in model-building - for example, they enable terms like  $\phi^i_{123} \psi_i  \phi^j_{123} s_j (...)$, with LH fermions and singlets. Such terms contribute to the effective neutrino terms through extended seesaws but do not contribute to charged fermions masses. In \cite{Ivo7}, the effective $@$ state arises similarly to the previously discussed case (there is some contribution from type II seesaw which does not alter the structure of the effective term). The $\odot$ state arises exclusively through extended seesaw, proceeding from the $\phi^i_{123} \psi_i  \phi^j_{123} s_j (...)$ term through a seesaw similar to the linear seesaw: the other two terms involved are simple FS invariants $\bar{s}^i s_i (...)$ and $\bar{s}^i \psi_i (...)$. Therefore $\phi^i_{123} \psi_i  \phi^j_{123} s_j (...)$, $\bar{s}^j s_j (...)$ and $\bar{s}^j \psi_j (...)$ combine to produce the desired $\odot$ eigenstate, with the structure of the contraction $\phi^j_{123} s_j$ passing into $\nu_j$.
For the full superpotential and other details of the model refer to \cite{Ivo7} - there are also several noteworthy features in the charged fermions sectors not discussed here.

Finally, it is worthwhile to mention \cite{Hagedorn:2008bc, Hirsch:2009mx} as different examples of models combining extended seesaw and FSs (note that those examples do not fit within the framework discussed previously).


\section{Conclusion}

GUTs and FSs have the potential to explain some of the observed features of the fermion masses and mixing. Combining the seesaw mechanism and a FS broken by special aligned VEVs in a GUT context is particularly ambitious, but possible despite the inherent constraints. We reviewed an interesting framework \cite{Ivo1, Ivo3, Ivo5} that is very suitable for this challenge, and then presented a generalization of this framework to extended seesaw scenarios \cite{Ivo7}.
Having mentioned some other examples in the literature, we conclude that seesaw mechanisms (extended or not) can be very useful in obtaining neutrino mixing in the context of GUTs with FSs.


\begin{theacknowledgments}
The work of IdMV was supported by FCT under the grant SFRH/BPD/35919/2007.
The work of IdMV was partially supported by FCT through the projects
POCI/81919/2007, CERN/FP/83503/2008
and CFTP-FCT UNIT 777  which are partially funded through POCTI
(FEDER) and by the Marie Curie RTN MRTN-CT-2006-035505.
\end{theacknowledgments}



\bibliographystyle{aipproc}   

\bibliography{refs}

\end{document}